\documentclass{PoS}
\usepackage{graphicx}
\usepackage{amssymb,amsfonts,amsmath}

\def\lsim{\mathrel{\rlap{\lower3pt\hbox{\hskip1pt$\sim$}}
     \raise1pt\hbox{$<$}}} 
\def\gsim{\mathrel{\rlap{\lower3pt\hbox{\hskip1pt$\sim$}}
     \raise1pt\hbox{$>$}}} 
\newcommand{\beq}{\begin{equation}}
\newcommand{\eeq}{\end{equation}}
\newcommand{\bea}{\begin{eqnarray}}
\newcommand{\eea}{\end{eqnarray}}

\title{QCD transition temperature: full staggered result}

\ShortTitle{QCD transition temperature: full staggered result}

\author{
Szabolcs~Bors\'{a}nyi$^a$,
Zolt\'{a}n~Fodor$^{a,b,c}$\footnote{speaker},
Christian Hoelbling$^a$,
S\'{a}ndor~D.~Katz$^c$,
Stefan~Krieg$^{a,d}$,
Claudia Ratti$^a$
and
K\'alm\'an~K.~Szab\'o$^a$\\
$^a$Department of Physics, University of Wuppertal, Gau\ss str. 20, D-42119,
Germany\\
$^b$Forschungszentrum J\"ulich, J\"ulich, D-52425, Germany\\
$^c$Institute for Theoretical Physics, E\"otv\"os University, P\'azm\'any
1, H-1117 Budapest, Hungary\\
$^d$Center for Theoretical Physics, MIT, Cambridge, MA 02139-4307, USA
}

\abstract{
We conclude our investigations on the QCD cross-over transition
temperatures with 2+1 staggered flavours and one-link stout improvement.  We
extend our previous two studies [Phys. Lett. B643 (2006) 46, JHEP 0906:088
(2009)] by choosing  even finer lattices ($N_t$=16) and we work again with
physical quark masses. These new results [for details see 
JHEP 1009:073,2010] support our earlier findings. 
We compare them with the published
results of the hotQCD collaboration.  All these results are confronted with the
predictions of the Hadron Resonance Gas model and Chiral Perturbation Theory
for temperatures below the transition region. Our results can be reproduced by
using the physical spectrum. The findings of the
hotQCD collaboration can be recovered only by using a distorted spectrum. 
This analysis provides a simple explanation for the observed
discrepancy in the transition $T$ between our and the hotQCD
collaborations. 
}

\FullConference{The XXVIII International Symposium on Lattice Field Theory, Lattice2010\\
		June 14-19, 2010\\
		Villasimius, Italy}

\begin{document}
{\bf Introduction.}
One of the most interesting quantities that can be extracted from lattice
simulations is the transition temperature $T_c$ at which hadronic matter passes
to a deconfined phase. $T_c$ has been vastly debated over the last few years, due to the
disagreement on its value observed by different lattice
collaborations, which in some cases is as high as 20\% of the absolute value.
Indeed, the analysis of the hotQCD collaboration (performed with two different
improved staggered actions, asqtad and p4, and with physical strange
quark mass and somewhat larger than physical $u$ and $d$ quark masses, $m_s
/m_{u,d}$=10), indicates that the transition region lies in the range $T =
(185-195)$ MeV. Different observables lead to the same value of $T_c$ 
(for the latest published result and for references see \cite{Bazavov:2009zn}). 
The authors expect that $m_s/m_u=20$
yields about 5~MeV shift (towards the smaller values) in the $T$
dependence of the studied observables.
On the other hand, the results obtained
by our collaboration using the staggered stout action (with physical light and
strange quark masses, thus $m_s/m_{u,d}\simeq$28) are quite different: $T_c$ 
lies in the range 150-170 MeV, and it
changes with the observable used to define it \cite{6, 7}. This is not
surprising, since the transition is a cross-over \cite{8}: in this case it is
possible to speak about a transition region, in which different observables may
have their characteristic points at different $T$ values, and the
$T$ dependences of the various observables play a more important role
than any single $T_c$ value.  Unfortunately, the 25-30 MeV discrepancy was
observed between the two groups for the $T$ dependences of the various
observables, too.

A lot of effort has been invested, to clarify the
discrepancy between the results of the two collaborations. (Note, that
quite recently preliminary results were presented \cite{Bazavov:2010sb} and
the results of the hotQCD collaboration moved closer to our results. We
include some of these data in our comparisons.) In Refs.
\cite{6,7}, we emphasized the role of the proper continuum limit
with physical $m_s/m_{u,d}$, showing how the lack of them can distort the
result.  In \cite{Fodor:2007sy} we pointed out that the continuum limit can be
approached only if one reduces the unphysical pion splitting (the main
motivation of our choice of action).  An interesting application of these
observations was studied in \cite{Huovinen:2009yb}. These authors have
performed an analysis within the Hadron Resonance Gas model (HRG).  They show that, to
reproduce the lattice results for the asqtad and p4 actions of the hotQCD
collaboration, it is necessary to distort the resonance spectrum away from the
physical one in order to take into account the larger quark masses used in
these lattice calculations, as well as finite lattice spacing effects.  As we
will see, no such distortion is needed to describe our data, and the
discrepancy between the two collaborations has its roots in the above mentioned
lattice artifacts.  

From the lattice point of view, we
present our most recent results for several physical quantities: our previous
works \cite{6, 7} have been extended to an even smaller lattice spacing
(down to $a \lsim0.075$ fm at $T_c$), corresponding to $N_t$=16. 
We use physical light and strange quark masses: we fix them by reproducing
$f_K/m_\pi$ and $f_K/m_K$ and by this procedure \cite{7} we get $m_s /m_{u,d} = 28.15$.

First we give the details of our numerical simulations. 
Then we present the results of our simulations for different observables.  We also present some aspects of the Hadron Resonance Gas 
model and the comparison 
between lattice and HRG model results. Finally we conclude. 

{\bf Details of the lattice simulations.}
We use \cite{6,7} a tree-level Symanzik 
improved gauge, and a stout-improved staggered fermionic action (see Ref. \cite{Aoki:2005vt} for details).  
The stout-smearing is an important part of the framework, which  
reduces the taste violation.

In analogy with what we did in \cite{6,7}, we set the scale at the physical 
point by simulating at $T=0$ with physical quark masses \cite{7} and reproducing the kaon and pion masses and the kaon decay constant. This gives 
an  uncertainty of about 2\% in the scale setting, which propagates in the uncertainty in the 
determination of the $T$ values listed.

\begin{figure}
\hspace{-.8cm}
\begin{minipage}{.48\textwidth}\parbox{6cm}{
\scalebox{.68}{\includegraphics{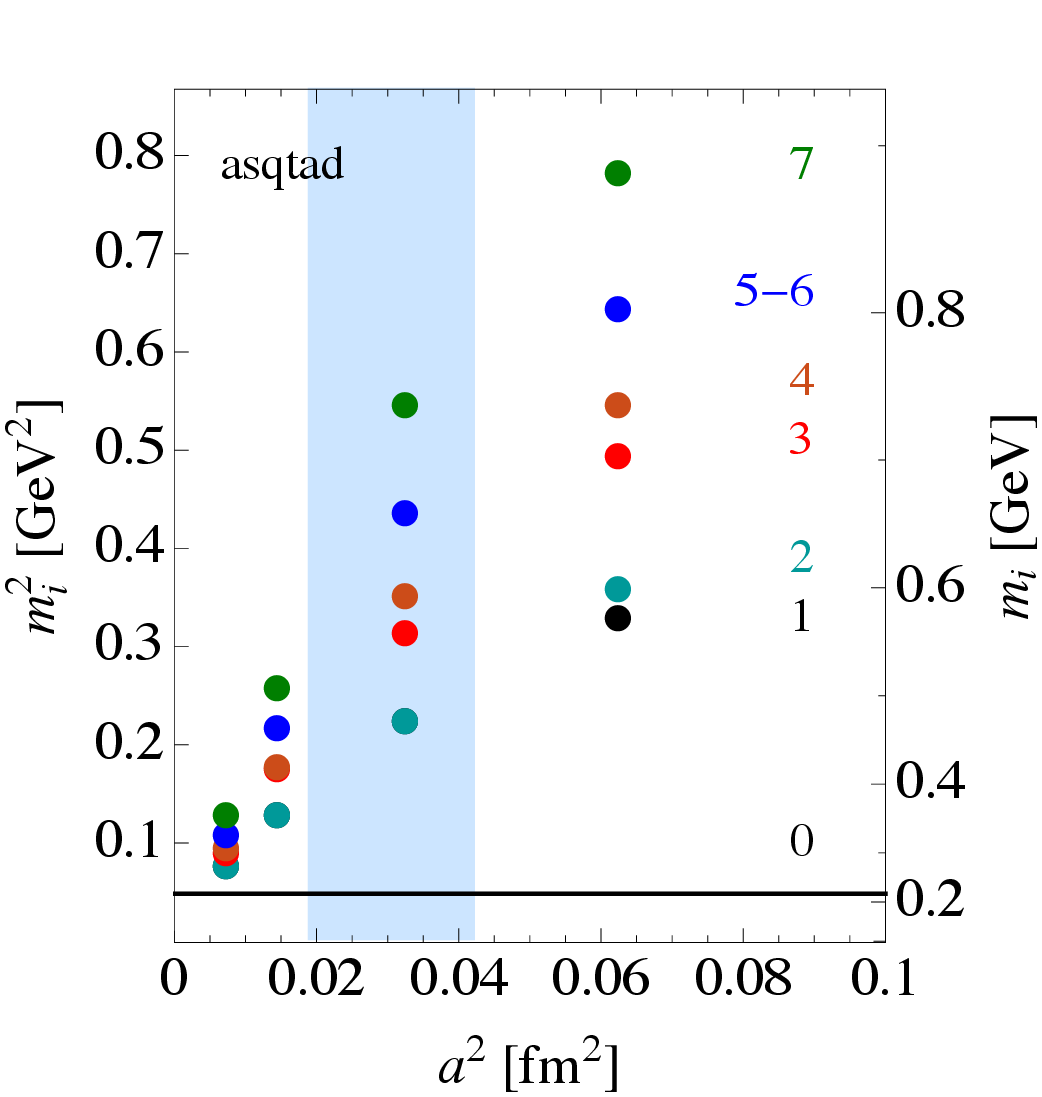}\\}}
%\centerline{(a)}
\end{minipage}
\hspace{.4cm}
\begin{minipage}{.48\textwidth}
\hspace{.4cm}
\parbox{6cm}{
\scalebox{.68}{
\includegraphics{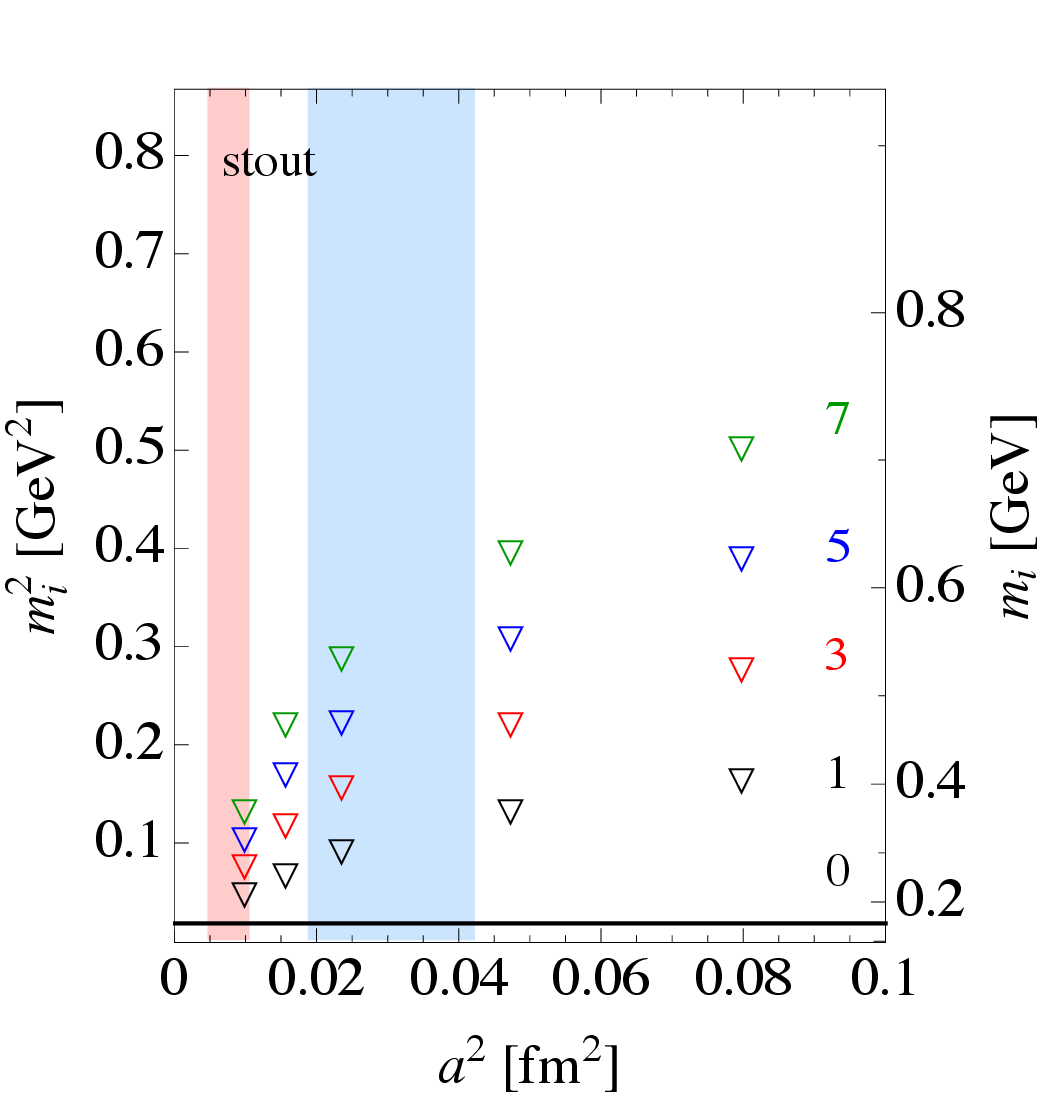}\\}}
%\centerline{(b)}
\end{minipage}
\caption{
Pion mass splitting, as functions of $a^2$. 
Left: asqtad action \cite{Bazavov:2009bb}. Right: stout action. 
In both panels, the blue band indicates the relevant range of
lattice spacings for a thermodynamics study at $N_t$=8 between $T$=120 and 180
MeV. The red band in the right panel corresponds to the same $T$ range
and $N_t$=16. 
}
\label{fig1}
\end{figure}

The pion splittings of a staggered framework 
are proportional to $(\alpha_s a^2 )$ for small $a$. 
It has to vanish in the continuum limit. Once it 
shows an $\alpha_s a^2$ dependence (in practice $a^2$ dependence
with a subdominant logarithmic correction) we are in the scaling region.  
This is an important 
check for the validity of the staggered framework at a given
lattice spacing. 
In Fig. \ref{fig1} we show the leading order $a^2$-behavior of the 
masses of the pion multiplets calculated with the asqtad (left) and stout (righ) actions. It is evident that the 
continuum expectation is reached faster in the stout action than in the asqtad one. In addition, 
in the present 
paper we push our results to $N_t = 16$, which corresponds to even smaller lattice 
spacings and mass splittings than those used in \cite{7}. 

%\FIGURE{

{\bf Lattice results.}
We present our lattice results for the strange quark number
susceptibility, Polyakov loop and two different definitions of the chiral
condensate. After performing a continuum extrapolation, we extract the values
of $T_c$ associated to these observables. 
The $T$ dependence of an observable
contains much more information than the location of a peak or
inflection point (which are usually hard to determine precisely for such a broad 
transition). 
We perform a HRG analysis and compare our results with those of the 
hotQCD Collaboration later. 

\begin{figure}
\hspace{-.8cm}
\begin{minipage}{.48\textwidth}
\parbox{6cm}{
\scalebox{.68}{
\includegraphics{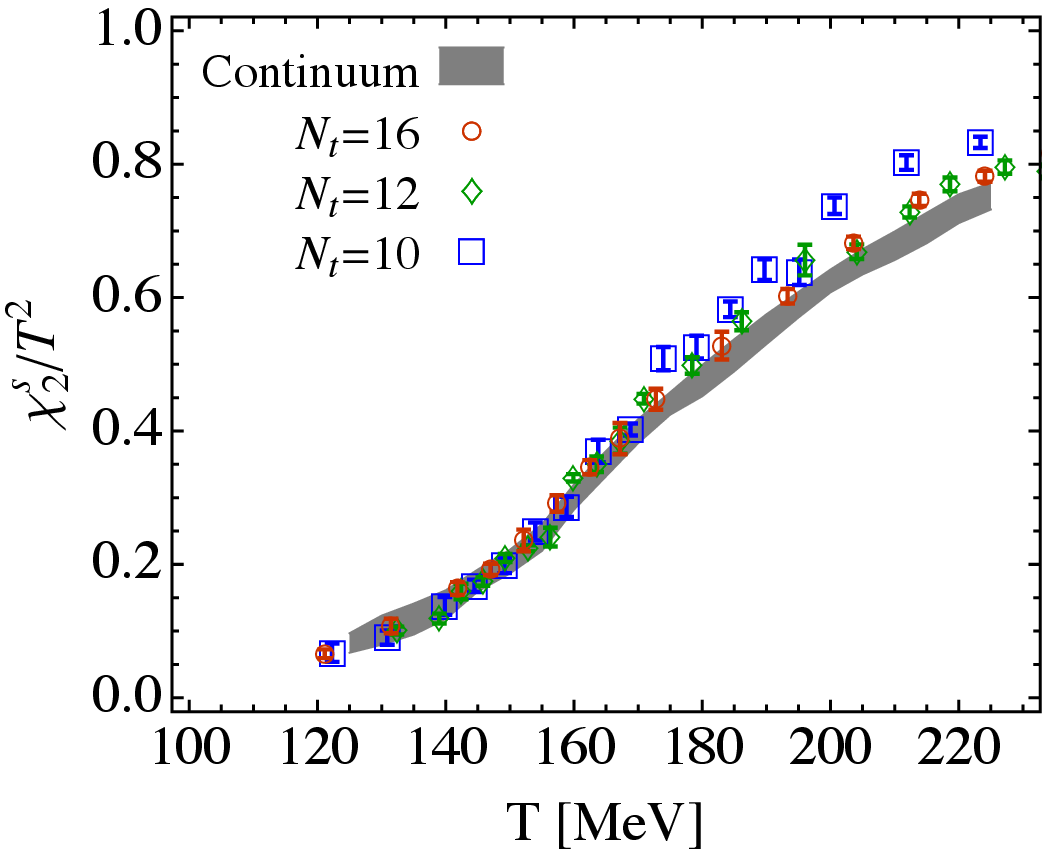}\\}}
%\centerline{(a)}
\end{minipage}
\hspace{.28cm}
\begin{minipage}{.48\textwidth}
\parbox{6cm}{
\scalebox{.68}{
\includegraphics{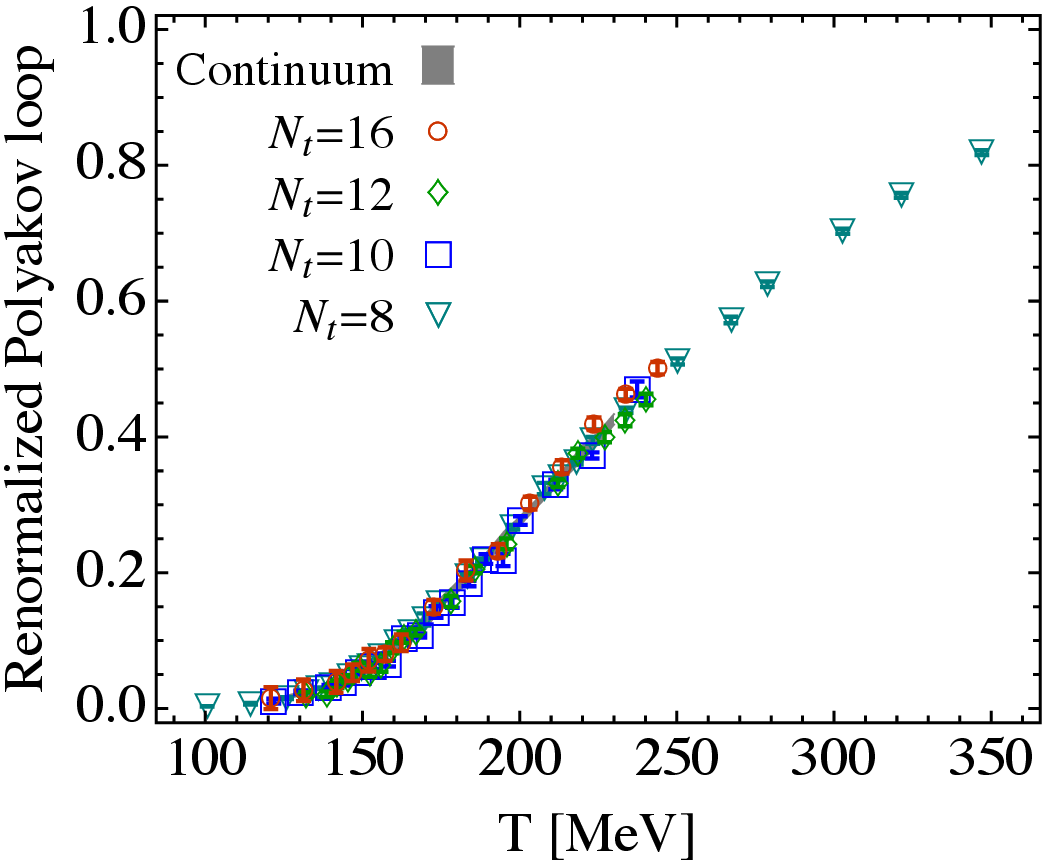}\\}}
%\centerline{(b)}
\end{minipage}
\caption{
Strange quark number susceptibility (left) and  Polyakov loop (right) as functions of $T$. }
\label{deco}
\end{figure}

Quark number susceptibilities 
increase during the transition, therefore
they can be used to identify this region. 
In the left panel of Fig. \ref{deco} we show our results for the strange quark
number susceptibility for $N_t=$ 10, 12, 16. The gray band shows our continuum
extrapolation.

The Polyakov loop indicates the
transition, since it exhibits a rise in the transition region.  In 
the right panel of Fig. \ref{deco} we plot the renormalized
Polyakov loop as a function of $T$. 
We use our
renormalization procedure of \cite{6}, in order to compare our results with
those obtained by the hotQCD collaboration \cite{Bazavov:2009zn} 
we use the same renormalization constant.
The various  $N_t$ data sets together with the
continuum extrapolated result are presented.
As it is expected from a broad cross-over the rise of the Polyakov
loop is pretty slow as we increase $T$ (c.f.
\cite{Bazavov:2009zn,6,7}).

%\FIGURE{

%\FIGURE{
\begin{figure}
\hspace{-.8cm}
\begin{minipage}{.48\textwidth}
\parbox{6cm}{
\scalebox{.68}{
\includegraphics{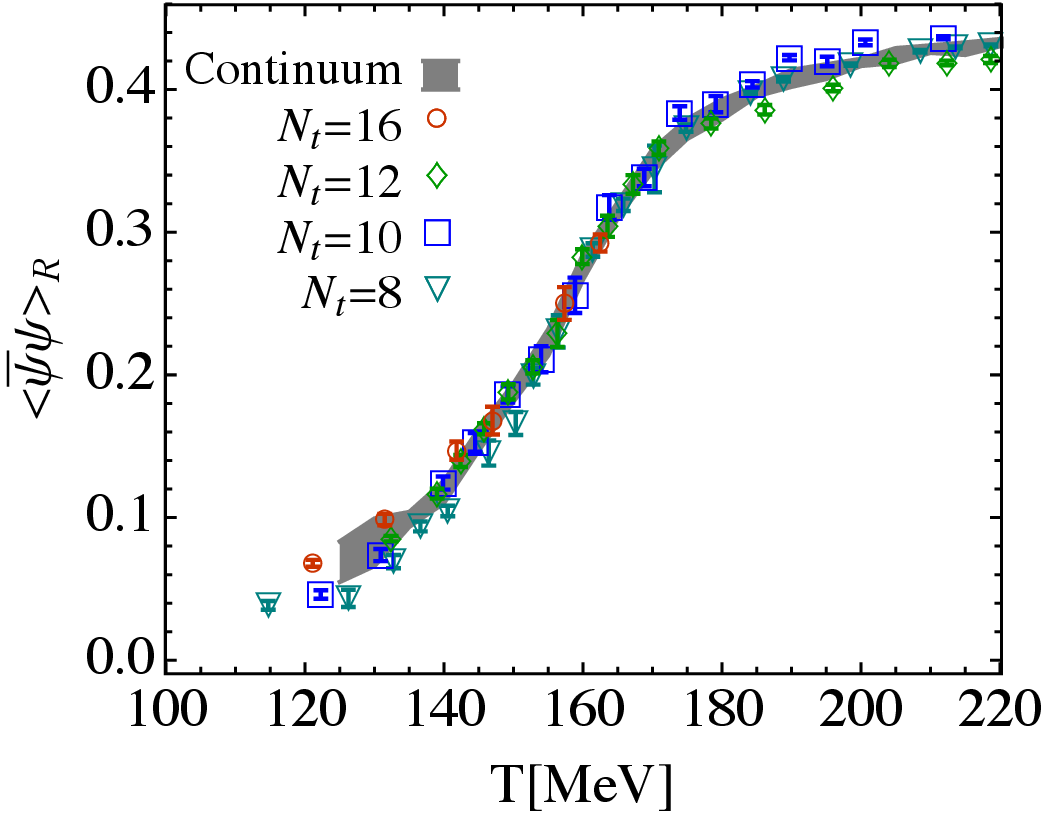}\\}}
%\centerline{(a)}
\end{minipage}
\hspace{.28cm}
\begin{minipage}{.48\textwidth}
\parbox{6cm}{
\scalebox{.68}{
\includegraphics{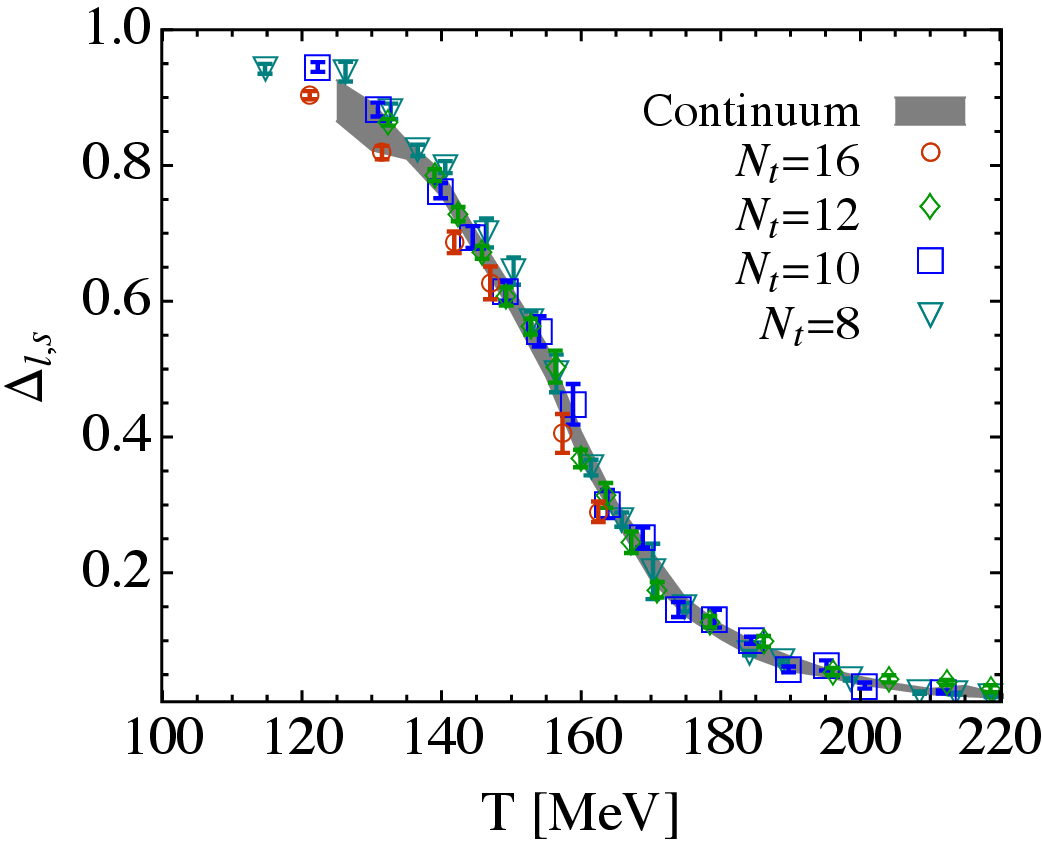}\\}}
%\centerline{(b)}
\end{minipage}
\caption{
Left: renormalized chiral condensate $\langle\bar{\psi}\psi\rangle_R$. 
Right: subtracted chiral condensate $\Delta_{l,s}$.  
}
\label{chir}
\end{figure}

The chiral condensate is defined as
$\langle\bar{\psi}\psi\rangle_q=T\partial\ln Z/(\partial m_q V)$ for q=u,d,s.
It can be
taken as an indicator for the remnant of the chiral transition, since it
rapidly changes around $T_c$.
We multiply the above expression by $m_q/m_\pi^4$ to define a dimensionless renormalized chiral condensate.
The individual results and the continuum extrapolation are shown 
in Figure~\ref{chir}.
In order to compare our results to those of the hotQCD collaboration, we also calculate the quantity
$\Delta_{l,s}$, which is defined as
$[{\langle\bar{\psi}\psi\rangle_{l,T}-{m_l}/{m_s}\langle\bar{\psi}\psi\rangle_{s,T}}]/
 [{\langle\bar{\psi}\psi\rangle_{l,0}-{m_l}/{m_s}\langle\bar{\psi}\psi\rangle_{s,0}}]$ for l=u,d.
Since the results at different lattice spacings are essentially 
on top of each other, we connect
them to lead the eye and use this band in later comparisons (c.f. 
Fig. \ref{chir}). 

{\bf Hadron Resonance Gas model} 
The HRG model has been widely used to study the
low $T$ phase of QCD in comparison with lattice data.
In Ref.
\cite{Huovinen:2009yb} an important ingredient was included,
the $m_\pi$- and $a$-dependence of the hadron masses \cite{19}. Here
we combine these ingredients with Chiral Perturbation Theory
($\chi$PT)  \cite{14}. This opens the possibility to study chiral quantities,
too.

The HRG model is based on the theorem of Ref. \cite{15}, 
which allows to calculate the microcanonical partition function of an interacting
system, for $V\rightarrow\infty$, 
to a good approximation, assuming that it is a gas of non-interacting free hadrons/resonances \cite{16}. 
The pressure of the model can be written as 
the sum of independent contributions coming from non-interacting resonances.
We include all known baryons and mesons up to 2.5 GeV, as listed in the latest edition of the PDG. 

We will compare the results obtained with the physical hadron masses to those 
obtained with the distorted one which takes into account $a$-effects. 
Each $\pi$/K in the staggered formulation is split into 16 mesons with different masses, which
are all included.
Similarly to Ref. \cite{Huovinen:2009yb}, we will also take into account the $m_\pi$- 
and $a$-dependence of all other hadrons/resonances. 

In order to calculate the chiral condensate in the HRG model, we need to know the behavior of all 
baryon and meson masses as functions of $m_l$ and $m_s$ . For  
the ground state hadrons we use 
\cite{MartinCamalich:2010fp}. 
The same study is not available for all the resonances that we include. Therefore, 
similarly to Ref. \cite{Huovinen:2009yb}, we work under the assumption that all resonance masses behave 
as their fundamental states as functions of $m_q$. In addition, 
we determine the contribution of pions to the chiral condensate obtained 
in three-loop $\chi$PT \cite{18}. 
All details of this calculation are given in \cite{Borsanyi:2010bp}. 

In our analysis we compare two sets of lattice data:\\
$\bullet$ {The first set is based on the Wuppertal-Budapest results.}\\
$\bullet$ {The second set is obtained by the Bielefeld-Brookhaven-Columbia-Riken
Collaboration, which later merged with a part of the the MILC collaboration 
and formed the hotQCD collaboration.} 

Furthermore, we use two types of 
theoretical descriptions (based on hadron resonance gas model and 
chiral perturbation theory, for short: HRG+$\chi$PT):\\
$\bullet$ { One of the theoretical descriptions is based on the physical spectrum from
 the PDG
 (we call this description ``physical").}\\
$\bullet$ { The other theoretical approach is based on a non-physical spectrum (this spectrum
 is obtained by $T=0$ simulations of the action one studies; the
 reason for this distortion will be explained later); we call this description ``distorted".}

As it is known, the Wuppertal-Budapest and the hotQCD results disagree. 
All characteristic $T$-s are higher for the hotQCD Collaboration.
Note, that this discrepancy is not related to the difficulty of determining
e.g. inflection points of slowly varying functions (typical for a broad
cross-over). The discrepancy appears for all variables for a large
$T$ interval. As we claimed earlier \cite{7} we observed 
``approximately 
20--35 MeV difference in the transition regime between our results and 
those of the hotQCD Collaboration".

As we will see, the Wuppertal-Budapest results are in complete agreement 
with the ``physical" HRG model and with the ``physical" 
chiral perturbation theory, whereas the hotQCD results cannot be 
described this way. The hotQCD results can only be described by the 
``distorted" HRG+$\chi$PT.

\begin{figure}
\hspace{-.8cm}
\begin{minipage}{.48\textwidth}
\parbox{6cm}{
\scalebox{.63}{
\includegraphics{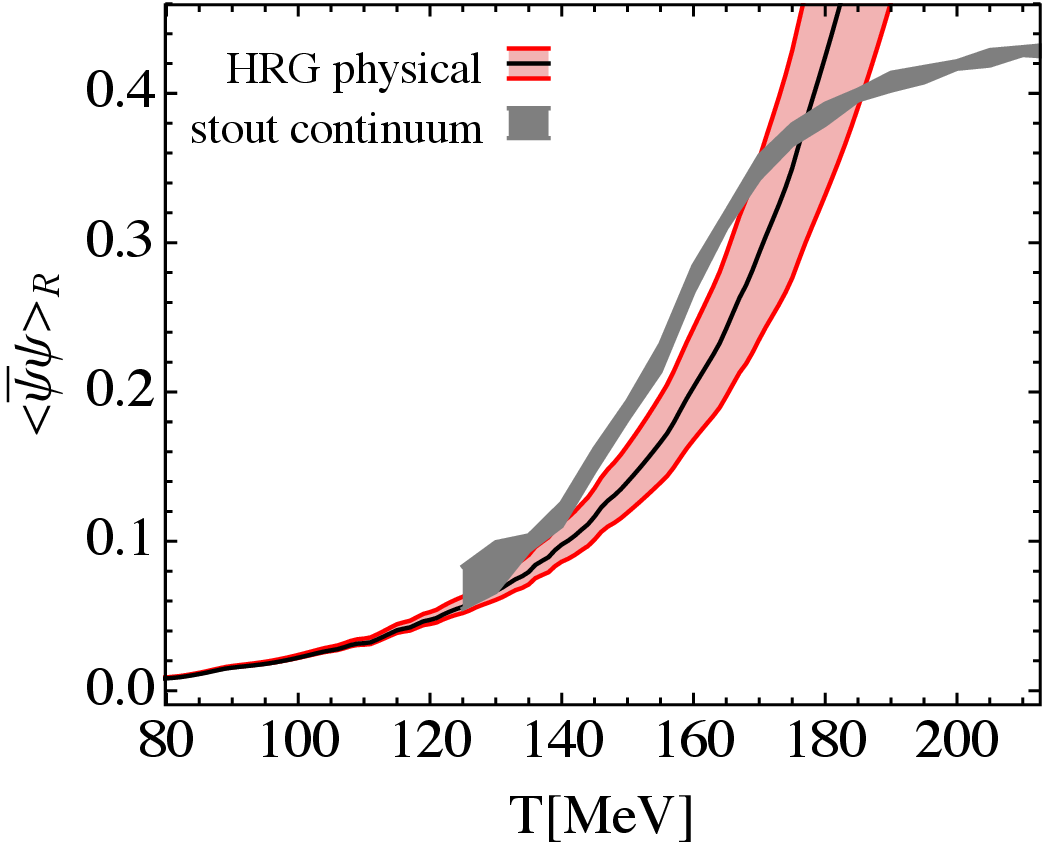}\\}}
%\centerline{(a)}
\end{minipage}
\hspace{.24cm}
\begin{minipage}{.48\textwidth}
\parbox{6cm}{
\scalebox{.63}{
\includegraphics{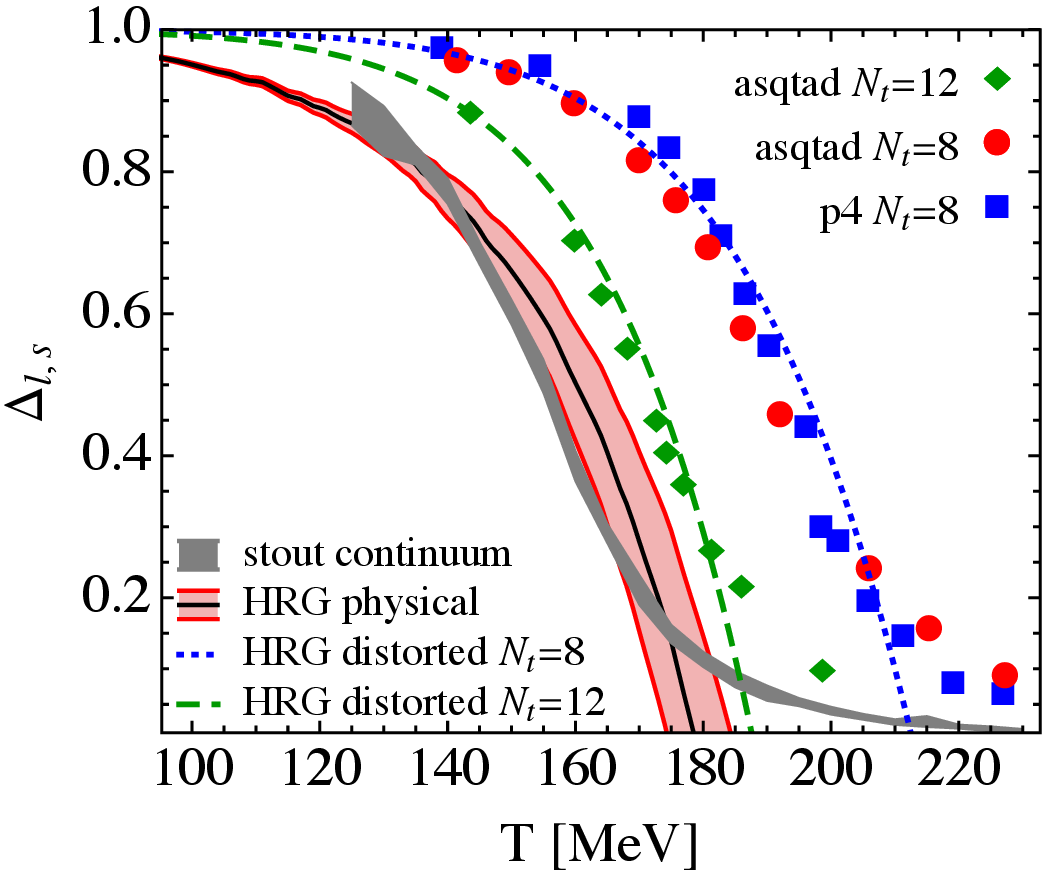}\\}}
%\centerline{(b)}
\end{minipage}
\caption{
Left: Renormalized chiral condensate. Right: $\Delta_{l,s}$. Both as a function of $T$. 
Gray bands are our continuum results, obtained with the 
stout action. Full symbols are obtained with the asqtad 
and p4 actions \cite{Bazavov:2009zn,Bazavov:2010sb}. In both panels, the solid line is the HRG model result with physical masses. 
The error band corresponds to the uncertainty in the quark mass-dependence of 
hadron masses. The dashed lines are the HRG+$\chi$PT model result with distorted masses of
the hotQCD Collaboration \cite{Bazavov:2009zn,Bazavov:2010sb} for $N_t=8$ and $N_t=12$.
}
\label{fig4}
\end{figure}

In Fig. \ref{fig4}, we show results for the chiral condensate as a function of
$T$.  The left panel shows $\langle\bar{\psi}\psi\rangle_R$, 
while the right panel shows $\Delta_{l,s}$.  
From all quantities that we have calculated, a consistent picture arises: our
stout results agree with the ``physical'' HRG+$\chi$PT predictions; whereas the
observed shift in $T_c$ between the results of the stout and
the asqtad and p4 actions can be easily explained within the Hadron Resonance
Gas+$\chi$PT model with ``distorted'' masses. Once the discretization effects,
the taste violation and the heavier quark masses used in
\cite{Bazavov:2009zn,Bazavov:2010sb} are taken into account, all the
HRG+$\chi$PT curves for the different physical observables are shifted to
higher $T$-s and fall on the corresponding lattice results. 

%\FIGURE{

\begin{figure}
\centerline{\includegraphics[width=2.8in]{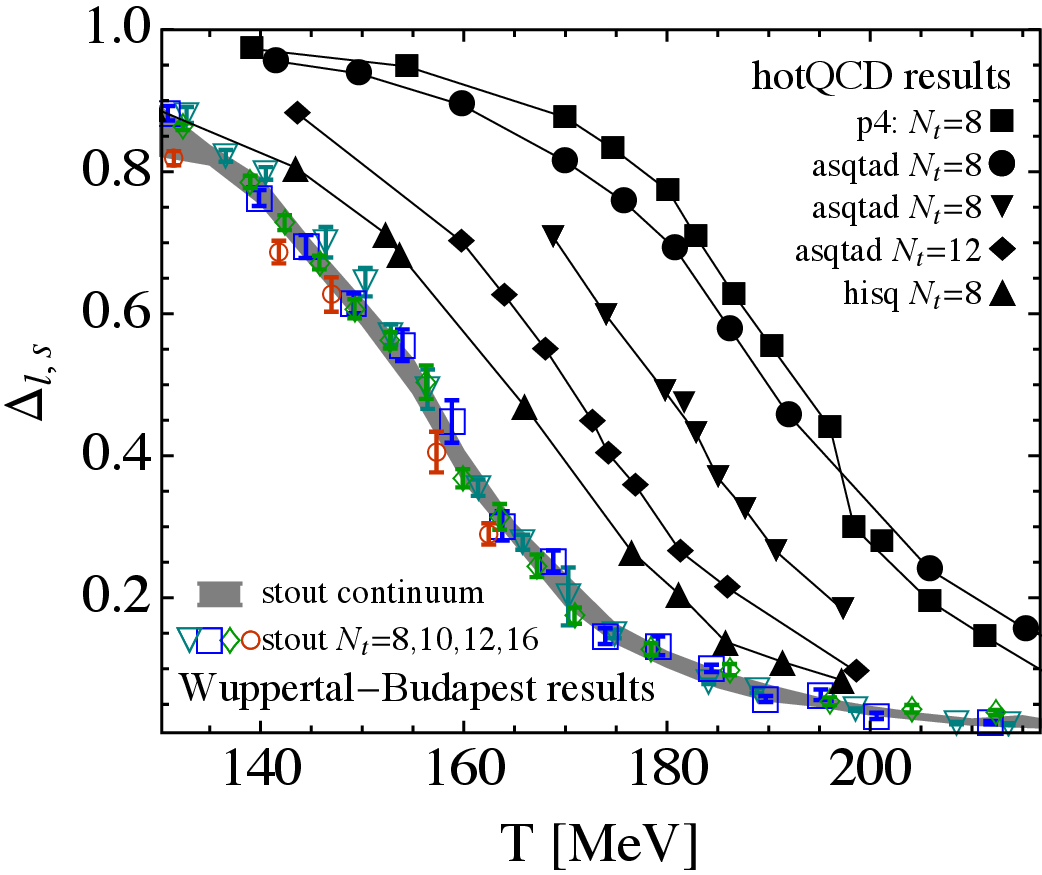}}
\caption{
$\Delta_{l,s}$ as a function of $T$.  We show a comparison between stout, asqtad, p4 and
HISQ \cite{Bazavov:2009zn,Bazavov:2010sb} results. 
Our stout results were all obtained by the
physical pion mass of 135 MeV. The full hotQCD dots and squares correspond to
$m_\pi=220$ MeV, the full triangles and diamonds correspond to $m_\pi=160$ MeV.  }
\label{fig5}
\end{figure}

As we mentioned there  are proceedings contributions written by two members of the hotQCD 
Collaboration, in which the HISQ action is applied and 
preliminary results are presented. 
The approximately 35 MeV discrepancy for the chiral condensate curves is 
reduced to about 10 MeV (see Fig. \ref{fig5}). 
Note, that the continuum limit within the HISQ 
framework is still missing. This last important step (which needs quite 
some computational resources and also care) will hopefully eliminate the 
remaining minor discrepancy, too. 
The same two members of the hotQCD
Collaboration presented preliminary results using the asqtad action on
$N_t$=12 lattices \cite{Bazavov:2010sb}, too. 
At this lattice spacing the pion splitting is smaller
than on $N_t$=8 lattices, and the curves move closer to ours. 
Following these authors (Figure 5. of Ref. \cite{Bazavov:2010sb}) we zoom in into the transition
region of $\Delta_{l,s}$ and on Figure~\ref{fig5}. 
The stout results from a broad range of "$a$"
($N_t$=8, 10, 12 and 16) are shown with open symbols. They are all 
in the vicinity of our continuum estimate, indicated by the thin gray band.
The hotQCD results were obtained by three different actions (p4, asqtad
and HISQ) and with two different pion masses (220 and 160~MeV). They
cover a broad range. The smaller the pion mass and/or splitting
in the hotQCD results, the closer it is to ours.

These confirm the expectations 
\cite{6,7} that the source of the discrepancy was 
the lack of the proper continuum extrapolation \cite{6} in the 
hotQCD result: a dominant discretization artefact within the asqtad 
and p4 actions is the large $\pi$ splitting \cite{Fodor:2007sy}, 
which resulted in the distorted spectrum. 

%\FIGURE{

{\bf Conclusions}
We have presented our latest results for the QCD transition temperature. The 
quantities that we have studied are the strange quark number susceptibility, the Polyakov loop,
the chiral condensate and the trace anomaly. We have given the complete
$T$ dependence of these quantities, which provide more
information than the characteristic $T$ values alone.
Our previous results for the strange quark susceptibility, the Polyakov loop and 
the chiral condensate have been pushed to an even finer lattice ($N_t$=16).
The new data corresponding 
to $N_t$=16 confirm our previous results. 
In order to find the origin of the discrepancy between the results of our collaboration and the hotQCD ones, we calculated these observables (except the Polyakov loop) in the Hadron 
Resonance Gas model. Besides using the physical hadron masses, we also
performed the calculation with modified masses which take into account the
heavier pions and larger lattice spacings used in \cite{Bazavov:2009zn}. We
find an agreement between our data and the HRG ones with ``physical'' masses,
while the hotQCD collaboration results are in agreement with the HRG model only
if the spectrum is ``distorted'' as it was directly measured on the lattice
\cite{Bazavov:2009bb}. This analysis therefore provides an easy and convincing
explanation of the observed shift in $T_c$ between the two
collaborations and emphasizes the role of the proper continuum limit.
All the details can be found in Ref. \cite{Borsanyi:2010bp}.


\begin{thebibliography}{99}

\bibitem{Bazavov:2009zn}
 A.~Bazavov {\it et al.},
  %``Equation of state and QCD transition at finite temperature,''
  Phys.\ Rev.\  D {\bf 80}, 014504 (2009)
  %[arXiv:0903.4379 [hep-lat]].
  %%CITATION = PHRVA,D80,014504;%%

\bibitem{6}
Y.~Aoki, Z.~Fodor, S.~D.~Katz and K.~K.~Szabo,
  %``The QCD transition temperature: Results with physical masses in the
  %continuum limit,''
  Phys.\ Lett.\  B {\bf 643}, 46 (2006)
  %[arXiv:hep-lat/0609068].
  %%CITATION = PHLTA,B643,46;%%
  
\bibitem{7}
Y.~Aoki, S.~Borsanyi, S.~Durr, Z.~Fodor, S.~D.~Katz, S.~Krieg and K.~K.~Szabo,
  %``The QCD transition temperature: results with physical masses in the
  %continuum limit II,''
  JHEP {\bf 0906}, 088 (2009)
  %[arXiv:0903.4155 [hep-lat]].
  %%CITATION = JHEPA,0906,088;%%
  
\bibitem{8}
Y.~Aoki, G.~Endrodi, Z.~Fodor, S.~D.~Katz and K.~K.~Szabo,
  %``The order of the quantum chromodynamics transition predicted by the
  %standard model of particle physics,''
  Nature {\bf 443}, 675 (2006)
  %[arXiv:hep-lat/0611014].
  %%CITATION = NATUA,443,675;%%
  
\bibitem{Bazavov:2010sb}
  A.~Bazavov and P.~Petreczky,
  %``Deconfinement and chiral transition with the highly improved staggered
  %quark (HISQ) action,''
  arXiv:1005.1131 [hep-lat]
  %%CITATION = ARXIV:1005.1131;%%

\bibitem{Fodor:2007sy}
  Z.~Fodor,
  %``QCD Thermodynamics,''
  PoS {\bf LAT2007} (2007) 011
  %[arXiv:0711.0336 [hep-lat]].
  %%CITATION = POSCI,LAT2007,011;%%

\bibitem{Huovinen:2009yb}
  P.~Huovinen and P.~Petreczky,
  %``QCD Equation of State and Hadron Resonance Gas,''
  Nucl.\ Phys.\  A {\bf 837} (2010) 26;
  %[arXiv:0912.2541 [hep-ph]].
  %%CITATION = NUPHA,A837,26;%%
  %P.~Huovinen and P.~Petreczky,
  %``On Fluctuations of Conserved Charges : Lattice Results Versus Hadron
  %Resonance Gas,''
  arXiv:1005.0324 [hep-ph]
  %%CITATION = ARXIV:1005.0324;%%

\bibitem{EOS}
S.~Borsanyi {\it et al.},
  %``The QCD equation of state with dynamical quarks,''
  arXiv:1007.2580 [hep-lat]
  %%CITATION = ARXIV:1007.2580;%%

\bibitem{Aoki:2005vt}
  Y.~Aoki, Z.~Fodor, S.~D.~Katz and K.~K.~Szabo,
  %``The equation of state in lattice QCD: With physical quark masses  towards
  %the continuum limit,''
  JHEP {\bf 0601}, 089 (2006)
  %[arXiv:hep-lat/0510084].
  %%CITATION = JHEPA,0601,089;%%
  
\bibitem{Bazavov:2009bb}
A.~Bazavov {\it et al.},
  %``Full nonperturbative QCD simulations with 2+1 flavors of improved staggered
  %quarks,''
  arXiv:0903.3598 [hep-lat]
  %%CITATION = ARXIV:0903.3598;%%
  
\bibitem{19}
S.~Durr {\it et al.},
  %``Ab-Initio Determination of Light Hadron Masses,''
  Science {\bf 322}, 1224 (2008)
  %[arXiv:0906.3599 [hep-lat]].
  %%CITATION = SCIEA,322,1224;%%
  
\bibitem{14}
J. Gasser and H. Leutwyler, Annals Phys. {\bf 158} (1984) 142

\bibitem{15}
R. Dashen, S. K. Ma and H. J. Bernstein, Phys. Rev. {\bf 187}, 345 (1969)

\bibitem{16}
R. Venugopalan and M. Prakash, Nucl. Phys. A {\bf 546}, 718 (1992).

\bibitem{MartinCamalich:2010fp}
  J.~Martin-Camalich, L.~S.~Geng and M.~J.~V.~Vacas,
  %``The lowest-lying baryon masses in covariant SU(3)-flavor chiral
  %perturbation theory,''
  arXiv:1003.1929 [hep-lat]
  %%CITATION = ARXIV:1003.1929;%%
  
\bibitem{18}
P. Gerber and H. Leutwyler, Nucl. Phys. B {\bf 321}, 387 (1989)

\bibitem{Borsanyi:2010bp}
  S.~Borsanyi et al.,
                  [Wuppertal-Budapest Collaboration],
  %``Is there still any Tc mystery in lattice QCD? Results with physical masses
  %in the continuum limit III,''
  JHEP {\bf 1009} (2010) 073
  %[arXiv:1005.3508 [hep-lat]].

\end{thebibliography}
\end{document}